\documentclass[letterpaper, 10 pt, conference]{ieeeconf} 

\IEEEoverridecommandlockouts                             
\usepackage{graphics} 
\usepackage{graphicx}
\usepackage{float}
\usepackage{epsfig} 
\usepackage{newtxmath} 
\usepackage{bbm} 
\usepackage{times} 
\usepackage{amsmath} 
\usepackage{hyperref}
\usepackage{color}
\usepackage{mathrsfs}

\def\bs{\boldsymbol}
\def\inr{\in\mathbb{R}}
\newcommand{\norm}[1]{\left\lVert#1\right\rVert}

\makeatletter
\let\NAT@parse\undefined
\makeatother

\usepackage[square, sort&compress, numbers]{natbib}

\usepackage{comment}

\def\real{\mathbb{R}}

\newcommand{\until}[1]{\{1,\dots, #1\}}

\newcommand\oprocendsymbol{\hbox{$\square$}}
\newcommand\oprocend{\relax\ifmmode\else\unskip\hfill\fi\oprocendsymbol}

\newtheorem{theorem}{Theorem}
\newtheorem{proposition}{Proposition}
\newtheorem{lemma}{Lemma}

\newtheorem{remark}{Remark}

\newtheorem{assumption}{Assumption}

\renewcommand{\theenumi}{\roman{enumi}}

\begin{document}

\title{In-flight actuator failure recovery of a hexrotor via multiple models and extended high-gain observers \thanks{This work has been supported in part by NSF Award IIS-1734272 and NASA MSGC Award NNX15AJ20H.}}

\author{Connor J. Boss and
        Vaibhav Srivastava
\thanks{C. J. Boss and V. Srivastava are with the Department of Electrical and Computer Engineering, Michigan State University, East Lansing, MI, 48823 USA e-mail: \texttt{\{bossconn, vaibhav\}@egr.msu.edu}}}

\maketitle

\begin{abstract}
We study an in-flight actuator failure recovery problem for a hexrotor UAV. The hexrotor may experience external disturbances and modeling error, which are accounted for in the control design and distinguished from an actuator failure. A failure of any one actuator occurs during flight and must be identified quickly and accurately. This is achieved through the use of a multiple-model, multiple extended high-gain observer (EHGO) based output feedback control strategy. The family of EHGOs are responsible for estimating states, disturbances, and are used to select the appropriate model based on the system dynamics after a failure has occurred. The proposed method is theoretically analyzed and validated through simulations and experiments.
\end{abstract}

\IEEEpeerreviewmaketitle

\section{Introduction}
With increased dependence on multi-rotor UAVs in many mission critical applications from infrastructure inspection to aerial cinematography to reconnaissance and surveillance, the demand for increasingly reliable vehicles is growing. In these applications, the loss of a vehicle poses significant threats to financial, security, or personnel interests. Improvements in control strategies as well as improvements in both software and hardware implementation have increased reliability greatly. However, there is still significant room for improvement in the face of actuator failures.


Actuator failure is of particular interest when it comes to reliability, as conventional multi-rotor UAVs will crash, or at least require an emergency landing, in the event of a failure. A failure on a quadrotor results in loss of control of one degree of freedom \cite{mueller2014stability}. A failure on an octorotor can result in dramatically reduced thrust capacity \cite{alwi2013fault,alwi2015sliding}. A standard hexrotor may not remain fully controllable in the event of a failure, see Appendix \ref{A:controllability}. However, if only a subset of failure possibilities are considered, recovery is possible \cite{vey2015structural}.

The main challenges in recovering from an actuator failure during flight are the ability to quickly detect the failure and to reconfigure the system while preserving stability. A complete actuator failure will cause a large rotational torque, which in turn causes the UAV to roll and pitch rapidly. If action is not taken extremely quickly the UAV can arrive at a configuration where it simply cannot recover.


The area of fault detection and isolation has been investigated for generalized systems \cite{hwang2009survey,boskovic1998stable,boskovic1999stable,chakravarty2017adaptive}, as well as for multi-rotor UAVs, including quadrotors \cite{mueller2014stability,xu2016fault,avant2018dynamics,dydek2012adaptive,lu2015active}, hexrotors\cite{vey2015structural,vey2016experimental}, and octorotors \cite{alwi2013fault,alwi2015sliding}. An $\mathcal{L}_1$ fault tolerant adaptive control approach is presented in \cite{xu2016fault} to handle partial loss of thrust on a quadrotor. This control however cannot overcome a complete actuator failure.

Recovering from an actuator failure on a hexrotor is a more challenging problem. The method \cite{vey2015structural,vey2016experimental} relies on a specific motor rotation sequence to enable the hexrotor to recover control when one of a certain subset of actuators fails. Another method \cite{voyles2012hexrotor} involves a special airframe design where the actuators are canted off plane to enable the actuators to apply forces in all six degrees of freedom. This method does support loss of any one of the actuators, however, given the orientation of the rotors, the configuration is not efficient for nominal flight. Since actuator failure is not a common occurrence, it is desirable to have a method for actuator failure recovery that does not compromise nominal flight performance and does not add weight or complexity. 

In this work, we study the problem of a hexrotor suffering a failure of any single actuator during flight and design a multiple-model, multiple EHGO output feedback control strategy to rapidly reconfigure the control of the hexrotor to regain stability after failure. We found that being able to rotate the rotors in either direction is necessary to maintain full controllability of the system under failure, see Appendix \ref{A:controllability}. This method is effective even in the presence of a wide range of modeling error and external disturbances. The proposed method is implemented algorithmically and verified in experiments. The contributions are as follows:
\begin{itemize}
    \item We devise a failure recovery strategy that enables recovery of any single actuator failure even in the presence of a broad class of disturbances.
    \item We rigorously analyze the proposed method to guarantee stability, guarantee correct model selection, and provide  bounds on the maximum model switching time following an actuator failure.
    \item We illustrate the effectiveness of the proposed approach through simulation and experimental results.
\end{itemize}

While preparing this manuscript, we became aware of the recent work~\cite{tzoumanikas2020nonlinear} which solves this problem using model predictive control. This work was completed independently, and is complimentary to \cite{tzoumanikas2020nonlinear}. However, we do use the method presented as a basis of comparison with our method.

The remainder of the paper is organized as follows. The system dynamics are introduced in Section \ref{sec:systemDynamics}, and the control law is designed in Section \ref{sec:outputFeedbackControlDesign}. The failure recovery strategy is described in Section \ref{sec:FailureRecoveryStrategy}, with stability analysis in Section \ref{sec:stabilityAnalysis}. Simulation results are presented in Section \ref{sec:simulation} with experimental results in Section \ref{sec:experiment} and conclusions in Section \ref{sec:conclusion}.

\section{System Dynamics}\label{sec:systemDynamics}
A hexrotor UAV is an underactuated mechanical system. To overcome the underactuation, the dynamics are split into two cascaded subsystems; the rotational dynamics and the translational dynamics.

\subsection{Rotational Dynamics}
Let $\bs{\theta}_1 = [\phi \ \theta \ \psi]^\top\in (-\frac{\pi}{2},\frac{\pi}{2})^2 \times (-\pi,\pi]$ be the Euler angles describing the hexrotor orientation in the inertial frame, and let $\bs{\theta}_2 = [\dot{\phi} \ \dot{\theta} \ \dot{\psi}]^\top \inr^3$ be the associated angular rates. Let $\bs{\theta}_r = [\phi_r \ \theta_r \ \psi_r]^\top \in (-\frac{\pi}{2},\frac{\pi}{2})^2 \times (-\pi,\pi]$ and $\dot{\bs{\theta}}_r = [\dot{\phi}_r \ \dot{\theta}_r \ \dot{\psi}_r]^\top \inr^3$ be the rotational reference signals. Define the rotational tracking error, $\bs \xi$, by
\begin{equation*}
    \begin{gathered}
        \bs{\xi}_1 = \bs{\theta}_1 - \bs{\theta}_r, \quad \bs \xi_2 = \dot{\bs{\xi}_1} = \bs{\theta}_2 - \dot{\bs{\theta}}_r, \quad \bs \xi = [\bs \xi_1^\top \; \bs \xi_2^\top]^\top. 
    \end{gathered}
\end{equation*}
Defining the inertia matrix, $J \inr^{3\times3}$, a matrix $\Psi \inr^{3\times3}$ which transforms body-fixed angular velocity to Euler angular rates \cite{boss2020ACC}, and its associated derivative, $\dot{\Psi}\inr^{3\times3}$, the rotational tracking error dynamics are

\begin{equation}\label{eq:rotationalErrorDynamicsFinal}
    \begin{split}
        \dot{\bs{\xi}}_1 &= \bs{\xi}_2, \\
        \dot{\bs{\xi}}_2 &= f(\bs{\xi},\bs{\theta}_1,\dot{\bar{\bs{\theta}}}_r) + G(\bs{\theta}_1)\bs{\tau} + \bs{\varsigma}_\xi,
    \end{split}
\end{equation}
where
\begin{equation*}
    \begin{split}
        f(\bs{\xi},\bs{\theta}_1,\dot{\bar{\bs{\theta}}}_r) &= \dot{\Psi}\Psi^{-1}(\bs{\xi}_2 + \dot{\bar{\bs{\theta}}}_r) \\
        &\quad -\Psi J^{-1}(\Psi^{-1}(\bs{\xi}_2 + \dot{\bar{\bs{\theta}}}_r) \times J\Psi^{-1}(\bs{\xi}_2 + \dot{\bar{\bs{\theta}}}_r)), \\
        G(\bs{\theta}_1) &= \Psi J^{-1},
    \end{split}
\end{equation*}
$\dot{\bar{\bs{\theta}}}_r$ is some approximation of $\dot{\bs \theta}_r$, $\bs{\tau}\inr^3$ is a vector of body-fixed torques, $\bs{\varsigma}_\xi = \bs{\sigma}_\xi - \ddot{\bs{\theta}}_r + [f(\bs{\xi},\bs{\theta}_1,\dot{\bs{\theta}}_r) - f(\bs{\xi},\bs{\theta}_1,\dot{\bar{\bs{\theta}}}_r)]\inr^3$ is an added term to represent the lumped rotational disturbance which satisfies \textit{Assumption \ref{as:disturbance}} (stated below), and $\bs{\sigma}_\xi \inr^3$ is the nominal rotational disturbance term \cite{boss2020ACC} in the original rotational dynamics with a generic control input.

\begin{assumption}[Properties of Disturbances]\label{as:disturbance}
For a control system with state $\bs x \in \real^n$, expressed in lower triangular form, such as~\eqref{eq:rotationalErrorDynamicsFinal}, any disturbance term is assumed to enter only the $x_n$ dynamics. The disturbance term  is also assumed to be continuously differentiable and its partial derivatives with respect to states are bounded on compact sets of those states for all $t\geq0$ \cite{boss2020ACC}.
\end{assumption}

\subsection{Translational Dynamics}
Let $\bs{p}_1 = [x \ y \ z]^\top\inr^3$ and $\bs{p}_2 = [\dot{x} \ \dot{y} \ \dot{z}]^\top\inr^3$ be the position and velocity of the hexrotor center of mass. Let $\bs{p}_r = [x_r \ y_r \ z_r]^\top \inr^3$ and $\dot{\bs{p}}_r = [\dot{x}_r \ \dot{y}_r \ \dot{z}_r]^\top \inr^3$ be the translational reference signals. 
Define the translational tracking error, $\bs \rho$, by
\begin{equation*}
    \begin{gathered}
        \bs{\rho}_1 = \bs{p}_1 - \bs{p}_r, \quad \bs{\rho}_2 = \dot{\bs{\rho}_1} = \bs{p}_2 - \dot{\bs{p}}_r, \quad \bs \rho=[\bs \rho_1^\top \; \bs \rho_2^\top]^\top.
    \end{gathered}
\end{equation*}
Taking the third column of the rotation matrix describing the hexrotor orientation in the inertial frame as $R_3(\bs{\theta}_1) \inr^3$, as in \cite{boss2020ACC}, $g$ as the gravitational constant, $u_f \inr$ as the total thrust input, $m\inr_{>0}$ as the mass, and defining $\bs{e}_z = [0 \; 0 \; 1]^T$, the translational tracking error dynamics are
\begin{equation}\label{eq:translationalErrorDynamics}
    \begin{split}
        \dot{\bs{\rho}}_1 &= \bs{\rho}_2, \\
        \dot{\bs{\rho}}_2 &= -\frac{u_f}{m}R_3(\bs{\theta}_1) + g\bs{e}_z + \bs{\sigma}_\rho - \ddot{\bs{p}}_r,
    \end{split}
\end{equation}
where $\bs{\sigma}_\rho \inr^3$ is an added term to represent the lumped translational disturbance which also satisfies \textit{Assumption \ref{as:disturbance}}.

\subsection{Failure Modes and Mapping Actuator Speeds to Inputs}
We will now consider how the system inputs in the form of body-torques, $\bs{\tau}$, and thrust force, $u_f$, are applied by the actuators, and how this changes during a failure. A thrust coefficient, $b \inr_{>0}$, relates rotor speed, $\omega \inr$, to force as
\begin{equation*}
    \bar{f}_j = b\omega_j^2\operatorname{sign}(\omega_j), \quad \text{for} \ j \in\{1,\dots,6\},
\end{equation*}
where $\bar{f}_j \inr$ can be positive or negative depending on the rotational direction of the rotor.
\begin{remark}[Bidirectional Rotor Rotation]\label{rem:bidirection}
    Bidirectional rotors are a requirement for a model switching failure recovery based on controllability analysis, see Appendix \ref{A:controllability}.
\end{remark} 

Define the failure matrix associated with a failure of rotor $i$ by $\mathcal{F}^{(i)} \inr^{6\times6}$ for $i \in\{0,\dots,6\}$, take $\mathcal{F}^{(0)} = I_6$, and
\begin{equation}\label{eq:failureMatrix}
        \mathcal{F}^{(i)} = \text{diag}(d_j),  \quad \text{with} \;  d_j = \begin{cases} 0, & \text{if} \; j = i, \\ 1, & \text{otherwise.}
        \end{cases}
\end{equation} 
Let $M \inr^{4\times6}$ be the mapping between actuator forces and system inputs and be defined by
\begin{equation}
    M = \left[\begin{smallmatrix} 1 & 1 & 1 & 1 & 1 & 1 \\ -\frac{r}{2} & -r & -\frac{r}{2} & \frac{r}{2} & r & \frac{r}{2} \\ \frac{r\sqrt{3}}{2} & 0 & -\frac{r\sqrt{3}}{2} & -\frac{r\sqrt{3}}{2} & 0 & \frac{r\sqrt{3}}{2} \\ c & -c & c & -c & c & -c \end{smallmatrix}\right],
\end{equation}
where $r \inr_{>0}$ is the distance from the hexrotor center of mass to the center of an actuator, and $c \inr_{>0}$ is the aerodynamic drag coefficient of a rotor. Let $i^*_t$ be the true configuration at time $t$, $i_t$ be the configuration that is selected at time $t$, and $t_f$ be the time of failure.

\begin{assumption}[Single Failure Occurrence]\label{as:singleFailure}
 We assume the configuration $i^*_t = i_t = 0$ for $t<t_f$. At the time of failure $i^*_{t_f} = 0 \rightarrow i^*_t \in \{1,\dots,6\}$ for $t > t_f$.
\end{assumption}

The system inputs for the nominal model and all failure models are mapped to a vector of directional squared rotor speeds, $\bs{\omega}_{s} = [\text{sign}(\omega_{1})\omega_{1}^2,\dots,\text{sign}(\omega_{6})\omega_{6}^2]^\top \inr^6$, through
\begin{equation}\label{eq:inputsToRotorSpeeds}
     \left[\begin{matrix} u_f \\ \bs{\tau} \end{matrix}\right]= b M\mathcal{F}^{(i_t)}\bs{\omega}_{s}.
\end{equation}

\section{Output Feedback Control Design} \label{sec:outputFeedbackControlDesign}
In this section, an output feedback estimation and control strategy is designed as in \cite{boss2020ACC}. We use a feedback linearizing control law and a family of EHGOs to estimate modeling error, external disturbances, and errors due to the failure of any one actuator. As such, each observer will correspond to a possible plant configuration, i.e., a nominal model and six failure models. The observer estimates will be used in the output feedback control, as well as for detecting and compensating for an actuator failure.

\subsection{Extended High-Gain Observer Design}\label{sec:EHGODesign}
A family of multi-input multi-output EHGOs is designed to estimate higher-order states of the error dynamic systems \eqref{eq:rotationalErrorDynamicsFinal} and \eqref{eq:translationalErrorDynamics}, and uncertainties arising from modeling error, external disturbances, and actuator failure \cite{khalil2017HGO}. It is shown in \cite{boss2017uncertainty,boss2020ACC} that it is necessary to include actuator dynamics in the multi-rotor model for EHGO design. 

The rotational and translational tracking error dynamics are combined and the state-space is extended to estimate disturbance vectors for both subsystems. For the purposes of writing the observer under the $i$-th failure mode, we write the extended system dynamics in the following form
\begin{equation}
    \begin{split}
    \dot{\bs{\rho}}_1 &= \bs{\rho}_2, \\
        \dot{\bs{\rho}}_2 &= -\frac{u_f}{m}R_3(\bs{\theta}_1) + g\bs{e}_z + \bs{\sigma}_\rho - \ddot{\bs{p}}_r, \\
        \dot{\bs{\sigma}}_\rho &= \varphi_\rho(t,\bs{\rho}),\\
        \dot{\bs{\xi}}_1 &= \bs{\xi}_2, \\
        \dot{\bs{\xi}}_2 &= f(\bs{\xi},\bs{\theta}_1,\dot{\bar{\bs{\theta}}}_r^{(i_t)}) + b\tilde{G}M\mathcal{F}^{(i)}\bs{\omega}_s^{(i_t)} + \bar{\bs{\varsigma}}_\xi^{(i)},\\
        \dot{\bar{\bs{\varsigma}}}_\xi^{(i)} &= \varphi_\xi^{(i)}(t,\bar{\bs{\varsigma}}_\xi^{(i)}),\\
    \end{split}
\end{equation}
where $\bar{\bs{\varsigma}}_\xi^{(i)} = \bs{\sigma}_m^{(i)} + \bs{\varsigma}_\xi$, and $\bs{\sigma}_m^{(i)} = b\tilde{G}M(\mathcal{F}^{(i^*_t)} - \mathcal{F}^{(i)})\bs{\omega}_{s}^{(i_t)}$ is the error resulting from an incorrect model, $i_t \neq i^*_t$. Finally, $\tilde{G} = [0_{3\times1} \; G(\bs{\theta}_1)]$ and $\bs{\sigma}_m^{(i^*_t)}=0$. Take $\varphi_\rho(t,\bs{\rho})$ and $\varphi_\xi^{(i)}(t,\bar{\bs{\varsigma}}_\xi^{(i)})$ as unknown functions describing the translational and rotational disturbance dynamics.
\begin{assumption}[Disturbance Dynamics]\label{as:disturbanceDynamics}
    It is assumed that $\varphi_\rho(t,\bs{\rho})$ and $\varphi_\xi^{(i)}(t,\bar{\bs{\varsigma}}_\xi^{(i)})$ are continuous and bounded on any compact set in the domain of $\bs{\rho}$ and $\bar{\bs{\varsigma}}_\xi^{(i)}$, respectively.
\end{assumption}


The observer system is written with extended states and directional squared rotor speeds as the input, $\bs{\omega}_s^{(i_t)}$. Defining the following state vectors
\begin{equation*}
    \bs{\chi}_1 = [\bs{\rho}_1^\top \; \bs{\rho}_2^\top \; \bs{\sigma}_\rho^\top]^\top, \; \; \bs{\chi}_2 = [\bs{\xi}_1^\top \; \bs{\xi}_2^\top \; \bs{\varsigma}_\xi^\top]^\top, \; \; \bs{\chi} = [\bs{\chi}_1^\top \; \bs{\chi}_2^\top]^\top,
\end{equation*}
we can write the EHGOs compactly as
\resizebox{.98\linewidth}{!}{
  \begin{minipage}{\linewidth}
\begin{equation}\label{eq:EHGOs}
    \begin{split}
        \!\!\dot{\hat{\bs{\chi}}}^{(i)} &= A\hat{\bs{\chi}}^{(i)} + B\left[\bar{f}(\hat{\bs{\chi}}^{(i)},\bs{\theta}_1,\dot{\bar{\bs{\theta}}}_r^{(i_t)}) + \bar{G}^{(i)}(\bs{\theta}_1)\bs{\omega}_s^{(i_t)} \right]+ H\hat{\bs{\chi}}_e^{(i)}, \\
        \hat{\bs{\chi}}_e^{(i)} &= C(\bs{\chi} - \hat{\bs{\chi}}^{(i)}),
    \end{split}
\end{equation}
\end{minipage}}
where $\hat{\bs{\chi}}^{(i)} = \left[\hat{\bs{\rho}}_1^{(i)\top} \ \hat{\bs{\rho}}_2^{(i)\top} \ \hat{\bs{\sigma}}_\rho^{(i)\top} \ \hat{\bs{\xi}}_1^{(i)\top} \ \hat{\bs{\xi}}_2^{(i)\top} \ \hat{\bar{\bs{\varsigma}}}_\xi^{(i)\top}\right]^\top$ is the estimate of $\bs \chi$ under the model with failure $i$, and
\resizebox{.99\linewidth}{!}{
  \begin{minipage}{\linewidth}
  \begin{equation*}
    \begin{gathered}
    \!    A = \text{diag}\left(A_j\right), B = \text{diag}\left(B_j\right), 
        C = \text{diag}\left(C_j\right),  H = \text{diag}\left(H_j\right),
    \end{gathered}
\end{equation*}
\end{minipage}}
\begin{equation*}
    \begin{gathered}
        A_j = \left[\begin{smallmatrix} 0_3 & I_3 & 0_3 \\ 0_3 & 0_3 & I_3 \\ 0_3 & 0_3 & 0_3\end{smallmatrix}\right], \ B_j = \left[\begin{smallmatrix} 0_3\\ I_3 \\ 0_3\end{smallmatrix}\right], \ H_j = \left[\begin{smallmatrix} \alpha_1/\epsilon I_3 \\ \alpha_2/\epsilon^2 I_3 \\ \alpha_3/\epsilon^3 I_3 \end{smallmatrix}\right], \\
        C_j = \left[\begin{smallmatrix} I_3 & 0_3 & 0_3\end{smallmatrix}\right], \ \text{for} \ j \in \{1,2\}, 
    \end{gathered}
\end{equation*}
\begin{equation*}
    \bar{f}(\hat{\bs{\chi}}^{(i)},\bs{\theta}_1,\dot{\bar{\bs{\theta}}}_r^{(i_t)}) = \left[\begin{matrix} g\bs{e}_z - \ddot{\bs{p}}_r\\ f(\hat{\bs{\xi}}^{(i)},\bs{\theta}_1,\dot{\bar{\bs{\theta}}}_r^{(i_t)})\end{matrix}\right],
\end{equation*}
\begin{equation*}
    \bar{G}^{(i)}(\bs{\theta}_1) = b\left[\begin{matrix} \frac{-R_3(\bs{\theta}_1)}{m} & 0_{3} \\ 0_{3\times1} & G(\bs{\theta}_1) \end{matrix}\right]M\mathcal{F}^{(i)},
\end{equation*}
where $H$ is designed by choosing $\alpha_1, \alpha_2, \alpha_3 \inr_{>0}$ such that
\begin{equation}
    s^3 + \alpha_1 s^2 + \alpha_2  s + \alpha_3,
\end{equation}
is Hurwitz and $\epsilon \inr_{>0}$ is a tuning parameter that must be chosen small enough. All EHGO estimates must also be saturated outside a compact set of interest to avoid peaking, see \textit{Remark 1} in \cite{boss2020ACC}.

\subsection{Output Feedback Control}\label{sec:outputFeedbackControl}
The output feedback controllers are written using the estimates from the corresponding EHGO, $\hat{\bs{\chi}}^{(i)}$. The family of translational output feedback controllers, induced by rotational reference signals and total thrust, become
\begin{equation*}\label{eq:translationalOutputFeedback}
    \begin{split}
        \hat{\phi}_r^{(i)} &= \tan^{-1}\left( \dfrac{-\hat{f}_y^{(i)}}{\sqrt{(\hat{f}_x^{(i)})^2 + (\hat{f}_z^{(i)}-g)^2}} \right), \quad \hat{\psi}_r^{(i)} = 0, \\
        \hat{\theta}_r^{(i)} &= \tan^{-1}\left(\dfrac{\hat{f}_x^{(i)}}{\hat{f}_z^{(i)}-g}\right), \quad \hat{u}_{f d}^{(i)} = -\dfrac{m(\hat{f}_z^{(i)}-g)}{\cos{\hat{\phi}_r^{(i)}} \cos{\hat{\theta}_r^{(i)}}},
    \end{split}
\end{equation*}
where $\hat{\bs{f}}_t^{(i)} = [\hat{f}_x^{(i)}\; \hat{f}_y^{(i)}\; \hat{f}_z^{(i)}]^\top = -\gamma_1\hat{\bs{\rho}}_1^{(i)} - \gamma_2\hat{\bs{\rho}}_2^{(i)} - \hat{\bs{\sigma}}_\rho^{(i)} + \ddot{\bs{p}}_r$. The family of rotational output feedback controllers become
\begin{equation*}
    \hat{\bs{\tau}}_d^{(i)} = G^{-1}(\bs{\theta}_1)\left[\hat{\bs{f}}_r^{(i)} - f(\hat{\bs{\xi}}^{(i)},\bs{\theta}_1,\dot{\bar{\bs{\theta}}}_r^{(i_t)})\right],
\end{equation*}
where $\hat{\bs{f}}_r^{(i)} =  -\beta_1\hat{\bs{\xi}}_1^{(i)} - \beta_2\hat{\bs{\xi}}_2^{(i)} - \hat{\bar{\bs{\varsigma}}}_\xi^{(i)}$. Note that the rotational reference signal estimates
$( \hat{\phi}_r^{(i_t)}, \hat{\theta}_r^{(i_t)},  \hat{\psi}_r^{(i_t)})$
are used to estimate $\dot{\bar{\bs{\theta}}}_r^{(i_t)}$ in $\hat{\bs{\tau}}_d^{(i)}$ (see \cite{boss2020ACC} for details).
We then arrive at the family of commanded directional squared rotor speeds
\begin{equation}
    \bs{\omega}_{s}^{(i)} = \frac{1}{b}(M\mathcal{F}^{(i)})^\dagger\hat{\bs{u}}^{(i)}  , \quad \hat{\bs{u}}^{(i)} = \left[\begin{matrix} \hat{u}_{f d}^{(i)} \\ \hat{\bs{\tau}}_d^{(i)} \end{matrix}\right],
\end{equation}
where $(M\mathcal{F}^{(i)})^\dagger = (M\mathcal{F}^{(i)})^\top((M\mathcal{F}^{(i)})(M\mathcal{F}^{(i)})^\top)^{-1}$ is the minimum energy pseudo-inverse of $M\mathcal{F}^{(i)}$. 

The rotational closed-loop system under input $\bs{\omega}_s^{(i_t)}$ for any $i_t$ regardless of $i^*_t$ reduces to the following perturbed linear system, since the mismatch is captured by $\bar{\bs{\varsigma}}_\xi^{(i)}$ 
\begin{equation}\label{eq:failureClosedLoop}
    \dot{\bs{\xi}} = A_\xi \bs{\xi} + \epsilon B_1 \bs{\delta}^{(i)}, \; A_\xi = \left[\begin{smallmatrix} 0_3 & I_3 \\ -\beta_1I_3 & -\beta_2I_3 \end{smallmatrix}\right], \; B_1 = \left[\begin{smallmatrix} 0_3 \\ I_3 \end{smallmatrix}\right],
\end{equation}
where
\begin{equation}\label{eq:epsilonDelta}
    \begin{gathered}
        \epsilon\bs{\delta}^{(i)} = \bs{\varsigma}_\xi + \bs{\sigma}_m^{(i)} - \hat{\bar{\bs{\varsigma}}}_\xi^{(i)} + \beta_1(\bs{\xi}_1 - \hat{\bs{\xi}}_1^{(i)}) + \beta_2(\bs{\xi}_2 - \hat{\bs{\xi}}_2^{(i)})\\
        \quad \quad \quad + \; \Delta f^{(i)}, \quad
        \Delta f^{(i)} = f(\bs{\xi},\bs{\theta}_1,\dot{\bar{\bs{\theta}}}_r^{(i_t)}) - f(\hat{\bs{\xi}}^{(i)},\bs{\theta}_1,\dot{\bar{\bs{\theta}}}_r^{(i_t)}).
    \end{gathered}
\end{equation}
The ability to write the mismatched closed-loop system as \eqref{eq:failureClosedLoop} means that if $\epsilon$ is chosen small enough, the system under output feedback will recover performance of the desired linear system, even in the presence of an actuator failure without requiring a model switch. 

\begin{remark}[Multiple Possible Recovery Strategies]\label{rem:multiple-recoveries}
     For the small constants, $\epsilon_1,\epsilon_2\inr_{>0}$, where $\epsilon_1 \ll \epsilon_2$, if we choose $\epsilon \in (0,\epsilon_1)$ a recovery can be achieved without requiring a model switch. If we choose $\epsilon \in (\epsilon_1,\epsilon_2)$, nominal disturbances can be compensated, however, the large disturbance, $\bs{\sigma}_m^{(i)}$, can result in large estimation error. Due to practical constraints on $\epsilon$ when it comes to implementation, such as sample rate and noise, we must choose $\epsilon \in (\epsilon_1,\epsilon_2)$. This motivates our use of multiple models and multiple observers for recovery (see Appendix \ref{A:Thm1P} for details).
\end{remark}

\begin{remark}[Domain of Operation]\label{rem:domainOfOperation}
We define the domain of operation as the region in which singularities are avoided in the feedback linearizing control design \cite{boss2020ACC}. Since $\bs{\sigma}_m^{(i)} = 0$ for $i = i^*_t$, for any single actuator failure, with $\epsilon \in (0,\epsilon_2)$ and $i_t = i^*_t$, the closed-loop rotational subsystem becomes the linearized system \eqref{eq:failureClosedLoop} with  a small perturbation $\epsilon\bs{\delta}^{(i)}$. Therefore, the domain of operation is the same for all $i_t = i^*_t$.
\end{remark}

\section{Failure Recovery Strategy}\label{sec:FailureRecoveryStrategy}
The most common external disturbances experienced by a multi-rotor during flight are aerodynamic (wind gusts, drag, etc.), and therefore primarily affect the translational dynamics, leaving little disturbance in the rotational dynamics during normal flight. However, during an actuator failure, a large rotational torque will be generated.
This large torque appears as a high magnitude disturbance, $\bs{\sigma}_m^{(i)}$, in the rotational dynamics, thus motivating the choice of monitoring the rotational subsystem for actuator failure detection.

For $\epsilon \in (\epsilon_1,\epsilon_2)$, when $i_t \neq i^*_t$ immediately after failure, the perturbation, $\epsilon\bs{\delta}^{(i_t)}$, is no longer small, and may make \eqref{eq:failureClosedLoop} leave the domain of operation. This behavior can be identified by monitoring an estimate of the derivative of a Lyapunov function for the rotational subsystem, using the method presented in \cite{freidovich2007lyapunov}. Therefore we can detect the failure, and then switch models to recover stability. Defining $t_s$ as the time of control switching, we can define a maximum switching time, $t_{s_\text{max}}$, such that $t_s < t_{s_\text{max}}$ ensures recovery from the failure before \eqref{eq:failureClosedLoop} leaves the domain of operation (see Appendix \ref{A:Thm1P} for an estimate of $t_{s_\text{max}}$).

\subsection{Estimating the Lyapunov Derivative}
Since the derivative of the Lyapunov function is not available, it will be estimated using estimates from the EHGOs, similar to \cite{freidovich2007lyapunov}. Following \textit{Assumption \ref{as:singleFailure}}, the system begins in the nominal operating regime, $i^*_t = 0$, therefore only the nominal Lyapunov function derivative must be estimated
\begin{equation}
    \dot{\hat{V}}_\xi = \hat{\bs{\xi}}^{(0)\top} P_\xi \dot{\hat{\bs{\xi}}}^{(0)} + \dot{\hat{\bs{\xi}}}^{(0)\top} P_\xi \hat{\bs{\xi}}^{(0)},
\end{equation}
where $P_\xi A_\xi + A_\xi^\top P_\xi = -I_6$. We use this estimate to test the following inequality to detect an actuator failure
\begin{equation}\label{eq:failureDetectorV}
    \dot{\hat{V}}_\xi \leq a_0 - \big\|\hat{\bs{\xi}}^{(0)}\big\|^2,
\end{equation}
where $a_0 \inr_{>0}$ is a small constant introduced to overcome the $O(\epsilon)$ estimation errors. Once \eqref{eq:failureDetectorV} is violated, a new model must be selected. 

\subsection{Estimating Disturbance and Failures Simultaneously}
Since all disturbance estimates contain any discrepancies between the modeled response and the response of the physical system, the total rotational disturbance estimated by the $i$-th observer, $\hat{\bar{\bs{\varsigma}}}_\xi^{(i)}$, is an estimate of $\bs{\sigma}_m^{(i)} + \bs{\varsigma}_\xi$. In order to select the appropriate model after a failure has been detected using \eqref{eq:failureDetectorV}, we utilize the magnitude of the disturbance estimates from each observer to appropriately select $i_t = i^*_t$ as
\begin{equation}\label{eq:modelSelect}
    i_t = \arg \min_{i \in\{1,\dots,6\}}\left\{\big\|{\hat{\bar{\bs{\varsigma}}}_\xi^{(i)}}\big\| \right\}.
\end{equation}
Following \textit{Assumption \ref{as:singleFailure}}, \eqref{eq:modelSelect} is a minimization across failure modes, excluding the nominal model.



\section{Stability Analysis}\label{sec:stabilityAnalysis}
Following the stability arguments in \cite{boss2020ACC} and the Theorems therein, the proposed output feedback control design presented here meets the same stability guarantees. We can show that these stability guarantees are also met under actuator failure without switching models when $\epsilon \in (0,\epsilon_1)$, and also hold for $\epsilon \in (\epsilon_1,\epsilon_2)$ so long as $i_t = i^*_t$.

We must now investigate the stability of the system during an actuator failure. Define the scaled observer error for the rotational system, $\bs{\eta}^{(i)} = [\bs{\eta}_1^{(i)} \; \bs{\eta}_2^{(i)} \; \bs{\eta}_3^{(i)}]^\top\inr^{9}$, by
\begin{equation*}
    \bs{\eta}_1^{(i)} = \frac{\bs{\xi}_1 - \hat{\bs{\xi}}_1^{(i)}}{\epsilon^2}, \quad \bs{\eta}_2^{(i)} = \frac{\bs{\xi}_2 - \hat{\bs{\xi}}_2^{(i)}}{\epsilon}, \quad \bs{\eta}_3^{(i)} = \bs{\varsigma}_\xi + \bs{\sigma}_m^{(i)} - \hat{\bar{\varsigma}}_\xi^{(i)}.
\end{equation*}
The entire rotational output feedback closed-loop system can now be written as the singularly perturbed system
\begin{subequations}\label{eq:entireClosedLoop}
    \begin{align}
        \dot{\bs{\xi}} &= A_\xi \bs{\xi} + \epsilon B_1 \bs{\delta}^{(i)}, \label{eq:slowVariables} \\
        \epsilon\dot{\bs{\eta}}^{(i)} &= \Lambda\bs{\eta}^{(i)} + \epsilon\left(B_2 \tfrac{\Delta f^{(i)}}{\epsilon} + B_3(\varphi_\xi^{(i)}(t,\bar{\bs{\varsigma}}_\xi^{(i)}) + \dot{\bs{\sigma}}_m^{(i)})\right), \label{eq:fastVariables}
    \end{align}
\end{subequations}
where the system dynamics \eqref{eq:slowVariables} are the slow variables, the observer error \eqref{eq:fastVariables} are the fast variables, and
\begin{equation*}
     \quad \Lambda = \left[\begin{smallmatrix} -\alpha_1I_3 & I_3 & 0_3\\ -\alpha_2I_3 & 0_3 & I_3\\ -\alpha_3I_3 & 0_3 & 0_3 \end{smallmatrix}\right], \quad B_2 = \left[\begin{smallmatrix} 0_3 \\ I_3 \\ 0_3 \end{smallmatrix}\right], \quad B_3 = \left[\begin{smallmatrix} 0_3 \\ 0_3 \\ I_3 \end{smallmatrix}\right].
\end{equation*}
By \textit{Assumption \ref{as:disturbanceDynamics}}, $\varphi_\xi^{(i)}(t,\bar{\bs{\varsigma}}_\xi^{(i)})$ is continuous and bounded, and it can be shown that $\dot{\bs{\sigma}}_m^{(i)}$ is also continuous and bounded, so we can bound the sum as, $\varphi_\xi^{(i)}(t,\bar{\bs{\varsigma}}_\xi^{(i)}) + \dot{\bs{\sigma}}_m^{(i)} \leq \Delta_\text{max}^{(i)}$.

From \cite{boss2020ACC}, $\Delta f^{(i)}$ is Lipschitz in $\bs{\xi}$ over the domain of operation and $\Delta f^{(i)}$ can be bounded by $\norm{\Delta f^{(i)}} \leq \epsilon L_\eta \norm{\bs{\eta}^{(i)}}$, for the Lipschitz constant, $L_\eta$. We can write a Lyapunov function for the scaled observer error system \eqref{eq:fastVariables} as
\begin{equation}\label{eq:etaLyapunovFunction}
    V_\eta^{(i)} = (\bs{\eta}^{(i)})^\top P_\eta \bs{\eta}^{(i)}, \quad \text{where} \; P_\eta \Lambda + \Lambda^\top P_\eta = -I_9.
\end{equation}
\begin{lemma}[Bounds on Observer Error]\label{lem:boundsOnObserverError}
Let the observer error at the time of failure $t_f$ be $\bs{\eta}^{(i)}(t_f)$. Then, the observer error for $t>t_f$ can be bounded by
\begin{equation*}
\begin{gathered}
    \big\|\bs{\eta}^{(i)}(t) \big\| \leq \left(\left(\sqrt{ V_\eta^{(i)}(t_f)}-\epsilon\frac{\kappa}{c}\right)e^{-\frac{c(t - t_f)}{\epsilon}} + \epsilon\frac{\kappa}{c}\right)/\sqrt{\lambda_\text{min}(P_\eta)}, \\
    c = \left(\tfrac{1}{\lambda_\text{max}(P_\eta)} - \tfrac{\lambda_\text{max}(P_\eta)\epsilon L_\eta}{\lambda_\text{min}(P_\eta)} \right), \quad \kappa = \tfrac{\lambda_\text{max}(P_\eta) \Delta_\text{max}^{(i)}}{\sqrt{\lambda_\text{min}(P_\eta)}},
\end{gathered}
\end{equation*}
where $\lambda_\text{min}(\cdot)$ and $\lambda_\text{max}(\cdot)$ are the minimum and maximum eigenvalues of the argument, respectively.
\end{lemma}
\begin{proof}
    See Appendix \ref{A:Lem1P}.
\end{proof}

We can now write \eqref{eq:epsilonDelta} as
\begin{equation}
    \epsilon\bs{\delta}^{(i)} = \epsilon^2\beta_1(\bs{\eta}_1^{(i)}) + \epsilon\beta_2(\bs{\eta}_2^{(i)}) + \bs{\eta}_3^{(i)} + \Delta f^{(i)},
\end{equation}
which can be bounded in terms of observer error as
\begin{equation}\label{eq:closedLoopErrorBound}
    \epsilon\bs{\delta}^{(i)} \leq \bs{\delta}_\text{max}^{(i)}(t) = (\epsilon^2\beta_1 + \epsilon(\beta_2 + L_\eta) + 1) \big\|\bs{\eta}^{(i)}(t)\big\|.
\end{equation}

\textit{Lemma \ref{lem:boundsOnObserverError}} shows that  estimation error, $\bs{\eta}^{(i)}$, converges to an $O(\epsilon \Delta_\text{max}^{(i)})$ neighborhood of the origin within $O(\epsilon)$ time. 
Actuator failure is significantly more dynamic than external disturbance, i.e., 
$\dot{\bs{\varsigma}}_\xi$ 
is relatively small as compared with $\dot{\sigma}_m^{(i)}$. Thus, $\Delta_\text{max}^{(i^*_t)} \ll \Delta_\text{max}^{(i)}$ for $i \ne i^*_t$, since $\dot{\bs{\sigma}}_m^{(i^*_t)}=0$. Therefore, as stated in \textit{Remark~\ref{rem:multiple-recoveries}}, $\epsilon$ can be chosen larger for the correct model than for any incorrect model, motivating the use of multiple models to reduce the total system disturbance.

In order to select the appropriate model after failure, as long as $\bs{\sigma}_m^{(i)}$, $\bs{\varsigma}_\xi$, and $\big\|\bs{\eta}_3^{(i)}(t_s)\big\|$ satisfy, 
\begin{equation}\label{eq:disturbanceBound}
    \big\|\bs{\sigma}_m^{(i)}\| \geq 2 \big\|\bs{\varsigma}_\xi \big\| + \big\|\bs{\eta}_3^{(i)}(t_s) \big\| + \big\|\bs{\eta}_3^{(i^*_t)}(t_s) \big\| ,
\end{equation}
 for each $i \in \until{6}\setminus \{i_t^*\}$, the observer estimate, $\hat{\bar{\bs{\varsigma}}}_\xi^{(i^*_t)}$, will be the smallest in magnitude at $t_s$, therefore, \eqref{eq:modelSelect} will select the appropriate model.

\begin{proposition}[Correct Model Selection]\label{prop:modelSelection}
Under the control input, $\bs{\omega}_{s}^{(i_t)}$, the family of observers \eqref{eq:EHGOs} will produce disturbance estimates, $\hat{\bar{\bs{\varsigma}}}_\xi^{(i)}$, for $i \in \{0,\dots,6\}$. If $\bs{\sigma}_m^{(i)}$, $\bs{\varsigma}_\xi$, and $\big\|\bs{\eta}_3^{(i)}(t_s)\big\|$ satisfy \eqref{eq:disturbanceBound}, the estimate $\hat{\bar{\bs{\varsigma}}}_\xi^{(i^*_t)}$ will be the smallest in magnitude and \eqref{eq:modelSelect} will select the correct model.


\end{proposition}

\begin{proof}
    See Appendix \ref{A:Prp1P}.
\end{proof}

\begin{remark}[Minimum Switching Time]\label{rmk:switching} 
    At $t_f$, $\big\|\bs{\eta}_3^{(i)}(t_f)\big\|$ may be large, but will decay to an $O(\epsilon\Delta_\text{max}^{(i)})$ neighborhood of the origin in $O(\epsilon)$ time. Therefore, there is some $O(\epsilon)$ time we must wait to switch for the observer estimates to converge. 
    Furthermore, $\big\|\bs{\eta}_3^{(i^*_t)}(t)\big\|$ will decay to an $O(\epsilon\Delta_\text{max}^{(i^*_t)})$ neighborhood of the origin, further reducing $\big\|\hat{\bar{\bs{\varsigma}}}_\xi^{(i^*_t)}\big\|$ as compared with the other model estimates, since $\Delta_\text{max}^{(i^*_t)} \ll \Delta_\text{max}^{(i)}$.
\end{remark}

\begin{theorem}[Stability During Actuator Failure]
    Let the state of the system \eqref{eq:failureClosedLoop} at the time of failure $t_f$ be such that $V_\xi(t_f) < a$ for some sufficiently small $a>0$. Then, there exist 
    $\epsilon_1,\epsilon_2>0$, and maximum switching time, $t_{s_\text{max}}>t_f$, such that the state, $\bs{\xi}$, will remain within the domain of operation during the failure transient, and  will recover tracking performance
    \begin{enumerate}
        \item  after the transient, when $\epsilon \in (0,\epsilon_1)$; 
        \item  if the correct model, $i_t = i^*_t$, is selected before $t_{s_\text{max}}$, when $\epsilon \in (\epsilon_1,\epsilon_2)$.
    \end{enumerate}

    %
\end{theorem}
\begin{proof}
    See Appendix \ref{A:Thm1P}.
\end{proof}

\section{Simulation}\label{sec:simulation}
The proposed method is simulated for a hexrotor system tracking a trajectory that is sinusoidal in the $x$ and $y$ directions, while holding position in $z$, $\bs{p}_r = [\sin(t) \; 0.5\sin(t) \; 0]^\top$. The system is simulated in discrete-time at 100Hz to replicate the experimental system. While having only position and orientation measurements, with added noise, the hexrotor is able to track the reference trajectory, suffer an actuator failure at 10 seconds into flight, and recover to resume tracking the trajectory after switching controllers. To showcase the ability of this method to select the appropriate model even in the presence of large disturbances, the system is simulated with large external rotational disturbances, $\bs{\sigma}_\xi = 12[\sin(t) \; \cos(t) \; \sin(t)]^\top$, and translational disturbances, $\bs{\sigma}_\rho = [\sin(t) \; \cos(t) \; \sin(t)]^\top$.

\begin{figure}
    \centering
    \includegraphics[width=0.48\textwidth]{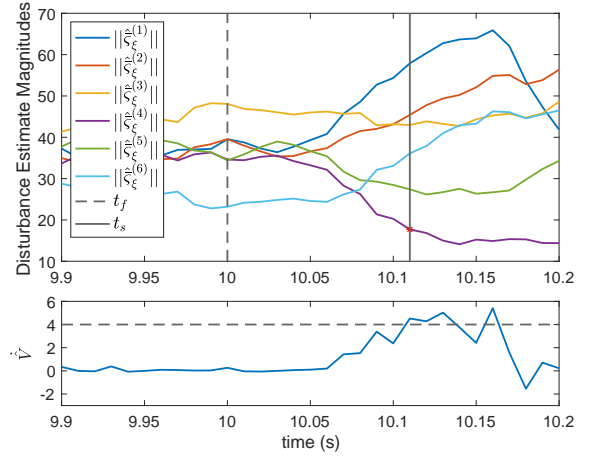}
    \caption{Norm of disturbance estimates for all failure case models and Lyapunov derivative estimate during a simulated in-flight actuator failure.}
    \label{fig:disturbanceMagnitude}
\end{figure}

The estimated Lyapunov function derivative, $\dot{\hat{V}}_\xi$, is monitored to determine when the failure occurs. 
Once the failure is detected, the controller is switched according to \eqref{eq:modelSelect}. The magnitudes of the estimated disturbances for all six failure modes, as well as the Lyapunov derivative estimate, are shown in Fig. \ref{fig:disturbanceMagnitude}. It quickly becomes apparent that the estimate of $\hat{\bar{\bs{\varsigma}}}_\xi^{(4)}$ has the smallest magnitude, which is indicative of a failure of actuator four. The dashed vertical line in Fig. \ref{fig:disturbanceMagnitude} corresponds to the time when a failure is induced, $t_f$, and the solid vertical line shows when the switch occurs, $t_s$, using \eqref{eq:modelSelect}. 


\begin{figure}
    \centering
    \includegraphics[width = 0.48\textwidth]{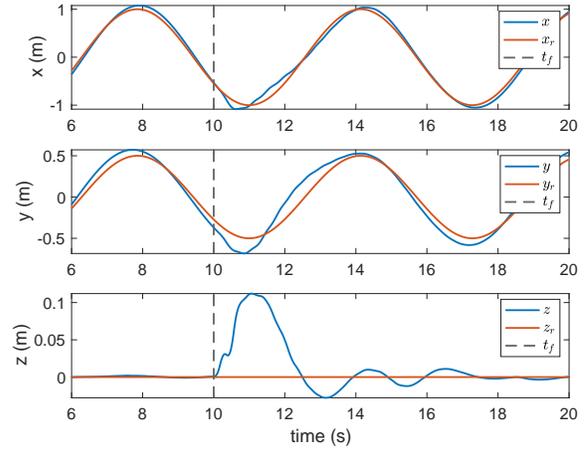}
    \caption{Hexrotor position tracking recovery after simulated in-flight actuator failure using multiple models to recover performance.}
    \label{fig:simulationResults}
\end{figure}

\begin{figure}
    \centering
    \includegraphics[width = 0.48\textwidth]{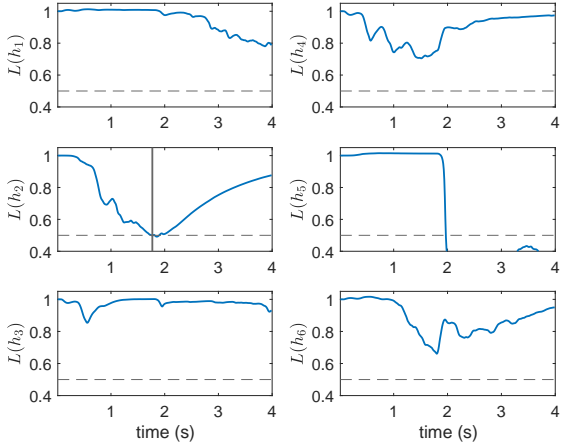}
    \caption{Health estimate of each rotor as estimated using the Extended Kalman Filter method.}
    \label{fig:rotorHealth}
\end{figure}




Fig. \ref{fig:simulationResults} shows the hexrotor briefly breaking from tracking the reference trajectory as the failure occurs at $t_f$, and after the controller is switched, the hexrotor successfully resumes tracking the trajectory. The tracking performance is slightly degraded due to the large external disturbances present in this simulation. 
For a hexrotor system this switch must occur rapidly to ensure the system can be re-stabilized by the newly selected controller.

To showcase the performance of our proposed algorithm, we compare with a rotor health estimation approach \cite{tzoumanikas2020nonlinear}. This method utilizes an Extended Kalman Filter which estimates the health of each rotor, $h_j$, for $j\in\{1,...6\}$. A logistic function is used to saturate the health estimates between 0 and 1. Utilizing the same dynamics and a detection cutoff on the health of each rotor of 0.5, as was shown to work well experimentally in \cite{tzoumanikas2020nonlinear}, we simulate this method. The same flight parameters and disturbances are again applied to the system with the EHGO based failure detection method replaced by the EKF method. Since the EKF method does not consider disturbances, the health estimates are affected by the disturbances applied to the system. Fig. \ref{fig:rotorHealth} shows that a failure of actuator two is detected just under two seconds into flight, when no failure has occurred. When the failure of actuator four occurs at ten seconds into flight, a recovery is not possible because the system is already operating in a failure mode for actuator two. We investigated lowering the cutoff threshold for the failure detection, however, this results in longer detection times and can result in a crash for these conditions in simulation because the switch does not happen quickly enough. In the presence of these disturbances, our method achieves a recovery where the EKF method fails. 

\section{Experimental Validation}\label{sec:experiment}
The proposed multiple-model estimation and control method is implemented on an experimental platform to validate performance and show recovery from an actuator failure during flight. The experimental platform is built on a 550mm hexrotor frame with 920kV motors and 10x4.5 carbon fiber rotors. Six 35A electronic speed controllers (ESCs) with bidirectional capability are used and the system is powered by a 5000mAh 4s LiPo battery.

The control method is implemented on a Pixhawk 4 FMU in discrete time at 100Hz using Mathworks Simulink through the \textit{PX4 Autopilots Support from Embedded Coder} package. 
We can access fused estimates of the vehicle orientation from the EKF running in the PX4 firmware. A Vicon motion capture system is used to provide indoor localization. 

Once the failure is detected and the controller is switched, the rotor opposite the failed rotor will be commanded to reverse directions to apply a large downward force to counteract the roll and pitch errors. Once the system returns to level flight, using the pseudo-inverse to compute desired rotor speeds results in the opposite rotor being commanded to apply small forces in either direction, thus requiring the rotor to change directions rapidly. During experimental testing it became clear that the opposite rotor could not change directions quickly enough to stabilize the system. To restrict the opposite rotor to only generate downward force for a detected failure, $i$, we impose force constraints, $f^{(i,j)}_{\min},f^{(i,j)}_{\max} < 0$ for the opposite rotor, $j$, defined by 
\begin{equation*}
    j =\begin{cases} i+1, & 1\le i \le 3, \\
    i-3, & 4 \le i \le 6,
    \end{cases}
\end{equation*}
and $f^{(i,j)}_{\min},f^{(i,j)}_{\max} \geq 0$ for all remaining $j$. These constraints ensure only a single directional change will be commanded when the model is switched. Let $\bar{\bs{f}}^*$ be the solution to the following optimization problem with above discussed constraints under the selected model $i_t$
%
%
%
\begin{equation*}
\begin{split}
\underset{\bar{\bs{f}}}{\text{minimize}} & \quad \left(\norm{\bar{\bs{f}}}^2 + \lambda \norm{W_v(M\mathcal{F}^{(i_t)}\bar{\bs{f}} - \hat{\bs{u}}^{(i_t)})}^2\right)\\
 \text{subject to} & \quad    f^{(i,j)}_{\min} \leq \bar{f}^*_{i,j} \leq f^{(i,j)}_{\max}, 
\end{split}
\end{equation*}

\begin{figure}
    \centering
    \includegraphics[width=0.48\textwidth]{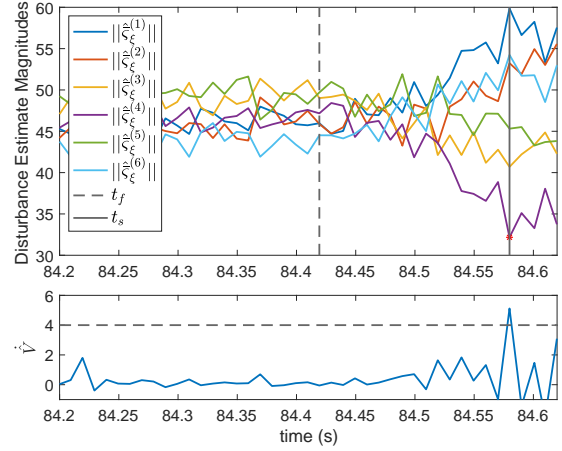}
    \caption{Norm of disturbance estimates for all failure case models and Lyapunov derivative estimate during an experimental in-flight actuator failure.}
    \label{fig:expDisturbanceNorms}
\end{figure}

\begin{figure}
    \centering
    \includegraphics[width=0.48\textwidth]{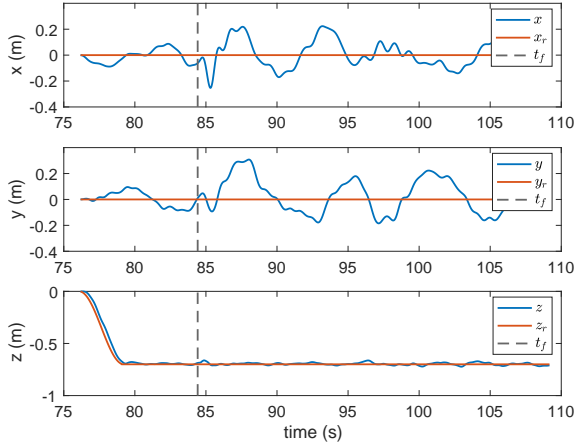}
    \caption{Hexrotor position tracking recovery after experimental in-flight actuator failure.}
    \label{fig:expTracking}
\end{figure}
\noindent
where $\lambda \inr_{>0}$ is chosen to be large to ensure we achieve applied forces and torques as close as possible to the desired $\hat{\bs{u}}^{(i_t)}$. We also take advantage of the diagonal weighting matrix, $W_v \inr^{4\times4}$, to prioritize the total thrust and the roll and pitch torques, allowing for lower performance in yaw tracking since the former are integral for achieving a successful recovery. The additional $\norm{\bar{\bs{f}}}^2$ term is used to simultaneously select a solution with lower energy, finally leading to the control $\bs{\omega}_{s}^{(i_t)} = \frac{1}{b}\bar{\bs{f}}^*$.
The optimization problem is solved by the active-set algorithm proposed in \cite{harkegard2002efficient}.

The experimental system is flown and a failure of actuator four is induced algorithmically. 
The norm of the experimental disturbance estimates for each failure model, and the Lyapunov derivative estimate, are shown in Fig. \ref{fig:expDisturbanceNorms}. 
The dashed vertical line in Fig. \ref{fig:expDisturbanceNorms} corresponds to the time when a failure is induced, $t_f$, and the solid vertical line shows when the detection and switch occurs, $t_s$, using \eqref{eq:modelSelect}.
The correct model for a failure of actuator four is selected and the resulting tracking performance before and after recovery are shown in Fig. \ref{fig:expTracking}. The tracking performance after failure is slightly degraded due to the use of non-ideal control inputs, $\bar{\bs{f}}^*$. A video of the experiments can be found at \url{https://youtu.be/1hJuWn53Rzo}

\section{Conclusion}\label{sec:conclusion}
We studied a trajectory tracking problem for a hexrotor in the presence of modeling error and external disturbances, while simultaneously enabling the hexrotor to recover stable flight and resume trajectory tracking in the presence of a complete actuator failure during flight. A multiple-model, multiple EHGO based output feedback control framework is used to enable this extended functionality. The framework is rigorously analyzed to provide stability guarantees and bounds on the maximum switching time for recovery. The approach is verified through simulation and shown to outperform an EKF based failure detection approach in the presence of large disturbances. Finally, experimental flight data shows that the proposed algorithm works on a physical system to recover from failure and resume stable flight.

\section{Acknowledgements}
We would like to thank Professor Hassan K. Khalil for his invaluable insights on extended high-gain observer design and analysis.

\appendix

\subsection{Proof of Lemma \ref{lem:boundsOnObserverError} (Bounds on Observer Error)}\label{A:Lem1P}
Taking the Lyapunov function \eqref{eq:etaLyapunovFunction} and computing the derivative with the scaled observer error system \eqref{eq:fastVariables} yields
\begin{equation*}
    \begin{split}
        \epsilon\dot{V}_\eta^{(i)} &= -(\bs{\eta}^{(i)})^\top\bs{\eta}^{(i)} \\
        &+ 2\epsilon(\bs{\eta}^{(i)})^\top P_\eta\left(B_2 \tfrac{\Delta f^{(i)}}{\epsilon} + B_3(\varphi_\xi^{(i)}(t,\bar{\bs{\varsigma}}_\xi^{(i)}) + \dot{\bs{\sigma}}_m^{(i)})\right),
    \end{split}
\end{equation*}
which can be bounded by
\begin{equation*}
    \begin{gathered}
        \epsilon\dot{V}_\eta^{(i)} \leq -\tfrac{c}{2} V_\eta^{(i)} + 2\epsilon\kappa \sqrt{V_\eta^{(i)}}, \\
        c = \left(\tfrac{1}{\lambda_\text{max}(P_\eta)} - \tfrac{\lambda_\text{max}(P_\eta)\epsilon L_\eta}{\lambda_\text{min}(P_\eta)} \right), \quad \kappa = \tfrac{\lambda_\text{max}(P_\eta) \Delta_\text{max}^{(i)}}{\sqrt{\lambda_\text{min}(P_\eta)}}.
    \end{gathered}
\end{equation*}
Taking $ W_\eta^{(i)} = \sqrt{V_\eta^{(i)}}$, the bound becomes
\begin{equation*}
    \dot{W}_\eta^{(i)} \leq -c W_\eta^{(i)} + \epsilon\kappa.
\end{equation*}
By the Comparison Lemma (Lemma 3.4 \cite{khalil2002nonlinear}), $W_\eta^{(i)}(t)$ is upper bounded by
\begin{equation*}
    W_\eta^{(i)}(t) \leq \left(W_\eta^{(i)}(t_f) - \epsilon\frac{\kappa}{c}\right)e^{-\frac{c}{\epsilon}(t-t_f)} + \epsilon\frac{\kappa}{c},
\end{equation*}
leading to the bound on scaled observer error
\begin{equation*}
     \big\|\bs{\eta}^{(i)}(t)\big\| \leq W_\eta^{(i)}(t)/\sqrt{\lambda_\text{min}(P_\eta)}.
\end{equation*}

\subsection{Proof of Proposition 1 (Correct Model Selection)}\label{A:Prp1P}
Suppose the modeling and external disturbances, $\bs{\varsigma}_\xi$, the disturbance resulting from incorrect model selection, $\bs{\sigma}_m^{(i)}$, and the scaled observer error, $\big\|\bs{\eta}_3^{(i)}(t_s)\big\|$ satisfy \eqref{eq:disturbanceBound}, then
\begin{align*}
        \hat{\bar{\bs{\varsigma}}}_\xi^{(i)} &=      \bs{\varsigma}_\xi + \bs{\sigma}_m^{(i)} - \bs{\eta}_3^{(i)} \\
        &\ge \|\bs{\sigma}_m^{(i)}\| -\big( \|\bs{\varsigma}_\xi\| + \|\bs{\eta}_3^{(i)}\|\big) \\
        &\underset{\text{using}~\eqref{eq:disturbanceBound}}{\ge} \big( \|\bs{\varsigma}_\xi\| + \|\bs{\eta}_3^{(i^*_t)}\|\big) \ge \bs{\varsigma}_\xi - \bs{\eta}_3^{(i_t^*)} = \hat{\bar{\bs{\varsigma}}}_\xi^{(i_t^*)},
\end{align*}
where the last equality holds since $\bs{\sigma}_m^{(i^*_t)}=0$.
Thus, the estimated disturbance, $\hat{\bar{\bs{\varsigma}}}_\xi^{(i^*_t)}$, will be the smallest in magnitude at $t_s$, and the solution to \eqref{eq:modelSelect} will be the correct model.

\subsection{Proof of Theorem 1 (Stability Under Actuator Failure)}\label{A:Thm1P}
A common Lyapunov function, $V_\xi$, for the feedback linearized rotational subsystem for each $i \in \{0,\dots,6\}$ and $\epsilon \to 0$ is given by
\begin{equation}\label{eq:V_xi}
    V_\xi = \bs{\xi}^\top P_\xi\bs{\xi}, \quad \text{where} \ P_\xi A_\xi + A_\xi^\top P_\xi = -I_6.
\end{equation}
Let $\Omega_\xi = \{V_\xi < c_\xi\}$  be an estimate of the domain of operation of the controller designed in Section~\ref{sec:outputFeedbackControl} for $c_\xi \inr_{>0}$ (see~\cite{boss2020ACC} for more details). For simplicity, we provide the estimate of the domain of operation for $\epsilon \to 0$, wherein the states and disturbances are estimated perfectly. For small $\epsilon > 0$, the obtained domain of operation can be shrunk to $\Omega'_\xi = \{V_\xi < c'_\xi\}$, with $c'_\xi < c_\xi$, to incorporate the effect of estimation error.

Using singular perturbation (\textit{Theorem 11.4} \cite{khalil2002nonlinear}) and non-vanishing perturbation (\textit{Lemma 9.2} \cite{khalil2002nonlinear}), it can be shown that \eqref{eq:entireClosedLoop} converges to an $O(\epsilon\Delta_\text{max}^{(i)})$ neighborhood of the origin for $\epsilon \in (0,\epsilon_2)$. For $\epsilon \in (0,\epsilon_1)$ the neighborhood $O(\epsilon\Delta_\text{max}^{(i)})$ is small enough for reasonable tracking performance. For $\epsilon \in (\epsilon_1,\epsilon_2)$, the large estimation error can make the trajectory leave the domain of operation and the system may diverge, thus requiring a model switch.

Taking the Lyapunov function \eqref{eq:V_xi}, and computing its derivative with the rotational closed-loop system \eqref{eq:failureClosedLoop} yields
\begin{equation}
    \dot{V}_\xi = -\bs{\xi}^\top \bs{\xi} + 2\bs{\xi}^\top P_\xi \epsilon B_1 \bs{\delta}^{(i)}.
\end{equation}
By the change of variables $W_\xi =\sqrt{V_\xi }$, and the arguments in Appendix \ref{A:Lem1P}, we can immediately upper bound $W_\xi(t)$ by
\begin{equation*}
    W_\xi(t_f)e^{{\frac{-(t-t_f)}{2\lambda_\text{max}(P_\xi)}}} + \int_{t_f}^t e^{{\frac{-(t-\tau)}{2\lambda_\text{max}(P_\xi)}}} \frac{\lambda_\text{max}(P_\xi)}{\sqrt{\lambda_\text{min}(P_\xi)}} \bs{\delta}_\text{max}^{(i)}(\tau) d\tau.
\end{equation*}
Let $t = t_{s_\text{max}}$ be the unique solution to the equation
\begin{equation*}
    \sqrt{a}e^{{\frac{-(t-t_f)}{2\lambda_\text{max}(P_\xi)}}} + \int_{t_f}^t e^{{\frac{-(t-\tau)}{2\lambda_\text{max}(P_\xi)}}} \frac{\lambda_\text{max}(P_\xi)}{\sqrt{\lambda_\text{min}(P_\xi)}} \bs{\delta}_\text{max}^{(i)}(\tau) d\tau = c_\xi.
\end{equation*}
The theorem follows immediately from the definition of $t_{s_{\max}}$. 

\subsection{Hexrotor Controllability Under Failure}\label{A:controllability}
In order to maintain controllablity of a hexrotor, five of the six actuators must remain functional. Simply disabling the rotor across the center of mass from the failed actuator and treating the system as a quadrotor does not result in a controllable system. This has to do with the rotational direction of the remaining rotors and becomes apparent by investigating the controllability of the linearized rotational subsystem about the hover equilibrium
\begin{equation}
    \dot{\bs{\theta}} = A\bs{\theta} + b B M\mathcal{F}^{(i)}\bs{\omega}_s,
\end{equation}
where $A = \left[\begin{smallmatrix} 0_3 & I_3 \\ 0_3 & 0_3 \end{smallmatrix}\right]$ and $\ B = \left[\begin{smallmatrix} 0_3 \\ J^{-1} \end{smallmatrix}\right]$. 
%
Here we will consider the failure matrix, $\mathcal{F}^{(i)}$, to include disabling the rotor across the center of mass as well 
\begin{equation*}
    \begin{split}
        \mathcal{F}^{(i)} &= \text{diag}(d_j),\\
        d_j &= \left\{
            \begin{array}{@{}ll@{}}
                0, & \text{if $j=i$ or $j=i+3$ when $1\leq i\leq 3$}, \\
                0, & \text{if $j=i$ or $j=i-3$ when $4 \leq i \leq 6$}, \\
                1, & \text{otherwise}.
            \end{array}\right.
    \end{split}
\end{equation*}
The controllability matrix for this system for $i\in\{1,\dots,6\}$ is not full rank, hence the system is not controllable.

Performing the same analysis for the linearized system with the failure matrix from \eqref{eq:failureMatrix}, the controllability matrix is full rank, hence the system is fully controllable with only five functional actuators. However, in order to achieve controllability, the actuators must be able to rotate in both directions to generate all necessary forces.

\footnotesize

\end{document}